\documentclass[aps,prc,reprint,groupedaddress]{revtex4-2}
\usepackage{amsmath}
\usepackage{amssymb}
\usepackage{wallpaper}
\usepackage{graphicx}
\usepackage{dcolumn}
\usepackage{bm}
\usepackage{color}
\usepackage{multirow}
\usepackage{CJK}
\usepackage{mathptmx}
\usepackage{tabularx}
\usepackage{tikz-feynman}
\usepackage{ulem}
\usepackage[T1]{fontenc}
\usepackage[colorlinks,linkcolor=blue,anchorcolor=blue,citecolor=blue]{hyperref}
\renewcommand{\emph}[1]{\textit{#1}}

\begin{document}
\begin{CJK*}{UTF8}{gbsn}
\renewcommand\arraystretch{1.2}
		
\title{Microscopic theory of the $\gamma$ decay of giant resonances in superfluid nuclei}

\author{W.-L. Lv
} 
\address{Frontiers Science Center for Rare isotopes, Lanzhou University, Lanzhou 730000, China}
\address{School of Nuclear Science and Technology, Lanzhou University, Lanzhou 730000, China}
\author{Y.-F. Niu
}\email{niuyf@sjtu.edu.cn} 
\address{School of Physics and Astronomy, Shanghai Jiao Tong University, 
    Key Laboratory for Particle Astrophysics and Cosmology (MoE), Shanghai 200240, China}
\address{Shanghai Key Laboratory for Particle Physics and Cosmology, Shanghai 200240, China}
\address{Frontiers Science Center for Rare isotopes, Lanzhou University, Lanzhou 730000, China}
\address{School of Nuclear Science and Technology, Lanzhou University, Lanzhou 730000, China}
\author{G. Col\`{o}}
\address{Dipartimento di Fisica, Universit\`{a} degli Studi di Milano, via Celoria 16, I-20133 Milano, Italy}
\address{INFN, Sezione di Milano, via Celoria 16, I-20133 Milano, Italy}

\date{\today}

\begin{abstract}
Recent advances in experiments have enabled the measurement of $\gamma$-decay 
from giant and pygmy resonances to low-lying states, establishing this technique
as a unique probe for nuclear structure.
However, a microscopic description of $\gamma$-decay to low-lying states in superfluid nuclei is still lacking.
We develop the Skyrme quasiparticle vibration (QPVC) model
to calculate $\gamma$-decay widths between vibrational states.
This model treats initial and final states as quasiparticle random phase approximation (QRPA) phonons 
and includes all the second-order diagrams for the interaction between the quasiparticles and the phonons,
while consistently accounting for the polarization processes. 
The same Skyrme functional is employed for the ground state and the interaction vertices.
As a timely application,
the $\gamma$-decay width from the giant dipole resonance (GDR) to the $2_{1}^{+}$ state in $^{140}$Ce is calculated,
which has recently been measured at the high intensity $\gamma$-ray source (HI$\gamma$S).
For the 4 Skyrme functionals we used, the total width of the collective dipole states in GDR region
is 250-420 eV and the corresponding branching ratio is 0.87-1.18\%.
The polarization effect, extracted microscopically,
agrees in trend with the macroscopic Bohr-Mottelson formula.
\end{abstract}

\maketitle
\end{CJK*}

\section{Introduction}
\label{secIntro}

Nuclear electromagnetic transitions serve as a fundamental probe of nuclear structure, 
offering crucial insights into the properties of excited states and the underlying many-body dynamics \cite{Alder1956}. 
For decades, they have played a central role in advancing the understanding of the nucleus, 
from revealing nuclear collective motion \cite{Glasmacher1998}, deformation \cite{Cline1986,Raman2001}, 
and new magic numbers \cite{Taniuchi2019}.
The emission of high-energy $\gamma$ rays \cite{Camera2023} serves 
as a damping mechanism for nuclear vibration modes \cite{Bertsch1983},
including pygmy and giant resonances.
Although the $\gamma$ decay contributes by a tiny fraction to the total damping, 
compared with spreading width and direct particle emission, 
it is particularly valuable because the electromagnetic interaction is well understood.
Precise $\gamma$-decay measurements therefore provide a clean probe of the underlying wave function 
\cite{Beene1989,Beene1990,Isaak2013,Wasilewska2022}.

The branching ratios of $\gamma$ decay to low-lying states are typically very small,
making such decays historically difficult to detect.
Recent experimental advances have begun to overcome these challenges.
Nuclear resonance fluorescence (NRF) experiments using 
the High Intensity $\gamma$-ray Source (HI$\gamma$S) \cite{Weller2009} 
enable detailed studies of $\gamma$ decays 
from dipole resonances to the $2_{1}^{+}$ state. 
In Ref. \cite{Kleemann2025}, the authors measured the $\gamma$ decay from 
the isovector giant dipole resonance (IVGDR) to the $2_{1}^{+}$ state in $^{154}$Sm. 
By comparing the measured branching ratio $\sigma_{2_{1}^{+}}/\sigma_{\rm ES}$ with 
macroscopic geometrical model predictions, 
they successfully extracted nuclear deformation parameters. 
In Ref. \cite{Papst2025}, a similar measurement for the pygmy dipole resonance (PDR) to the $2_{1}^{+}$ state in $^{150}$Nd revealed 
a significant deviation from the Porter-Thomas distribution, 
suggesting that the $\gamma$SF used in statistical models \cite{TALYS2023} may require revision. 
These findings demonstrate that high-energy $\gamma$ decays 
from excited resonances to low-lying states are a sensitive probe of nuclear structure 
and play an important role in refining nuclear astrophysics inputs.
More measurements will be carried out at 
the Laboratori Nazionali di Legnaro (LNL) \cite{LNL2023}
and Shanghai Synchrotron Radiation Facility (SSRF) \cite{Chen2023}.


Theoretically, earlier theoretical studies reporting $\gamma$-decay widths 
between giant resonances (GRs) and low-lying states 
have been performed mainly for closed-shell nuclei, most notably $^{208}$Pb, 
within the nuclear field theory (NFT) \cite{Bortignon1984,Bes1986},
extended theory of finite Fermi systems (ETFFS) \cite{Speth1985}, 
and the quasiparticle phonon model (QPM) \cite{Voronov1990}.
{The QPM has also been applied to open-shell nuclei \cite{Tsoneva2025}, 
in particular to the GDR decay in the $^{116,124}$Sn isotopes \cite{Ponomarev1992} 
and to the decay of low-energy dipole
excitations in $^{140}$Ce and $^{88}$Sr \cite{Loeher2016,Isaak2023}.}
These approaches incorporate phenomenological ingredients to different degrees.
In the NFT calculations, the strengths of schematic particle-hole
interactions are either adjusted using the energies of low-lying states and
giant resonances \cite{Bortignon1984,Bortignon1981}
or constrained by the empirical nucleon-nucleus optical potential \cite{Bes1986,Bes1975}. 
In the ETFFS, a parameterized particle-hole interaction is employed, 
with its parameters stated to be taken from experiment, 
while the required single-particle energies are either calculated 
or taken directly from experimental data \cite{Speth1985,Speth1970}.
In the QPM calculations, 
the residual interaction is modeled by a separable multipole force 
whose parameters are constrained by low-lying spectroscopic data
or the energies of GRs \cite{Tsoneva2025}.

Over the past few decades, nuclear density functional theory has been
widely applied to the description of medium- and heavy-mass nuclei \cite{Bender2003}. 
It provides a global description of nuclear properties across broad regions of the nuclear chart, 
and its extensions to nuclear dynamics have enabled microscopic studies of collective excitations, 
including giant resonances and rotational bands.
In recent years, the fully self-consistent treatment of $\gamma$ decay of GRs
with the Skyrme particle vibration coupling (PVC) model
has become available \cite{Brenna2012,Lv2021}.
In this model the dynamic correlations are considered through the PVC,
which has been found important for 
the single-particle levels \cite{Litvinova2006,Colo2010,Litvinova2011,Colo2017}, 
the Gamow-Teller response and the related $\beta$ decay \cite{Niu2014,Niu2015,Niu2018,Robin2019,Litvinova2020,Liu2024}, 
as well as the spreading widths and centroid energies of giant resonances \cite{RocaMaza2017,Shen2020,Li2023,Litvinova2023,Li2024}.
However, pairing correlations have never been included 
in the study of the $\gamma$ decay from GRs to low-lying states with the (Q)PVC model.

In this work, we extend this fully self-consistent approach to superfluid systems 
by the Skyrme QPVC model. 
The model incorporates pairing correlations based on Skyrme energy density functionals 
and is formulated to describe $\gamma$ decays between vibrational states within the NFT framework.
Motivated by recent experiments at the HI$\gamma$S \cite{Kleemann2024},
we employ our model to investigate the $\gamma$ decay of IVGDR to the $2_{1}^{+}$ state in $^{140}$Ce.
This paper is organized as follows. 
In Sec. \ref{secTheo}, we present our theoretical formalism. 
Section \ref{secNume} provides the numerical details of the calculation. 
Our results for $^{140}$Ce are presented and discussed in Sec. \ref{secResu}. 
We conclude with a summary in Sec. \ref{secSum}.

\section{Formalism}
\label{secTheo}

\begin{figure*}[tbph]
    \centering
    \includegraphics[width=0.9\textwidth]{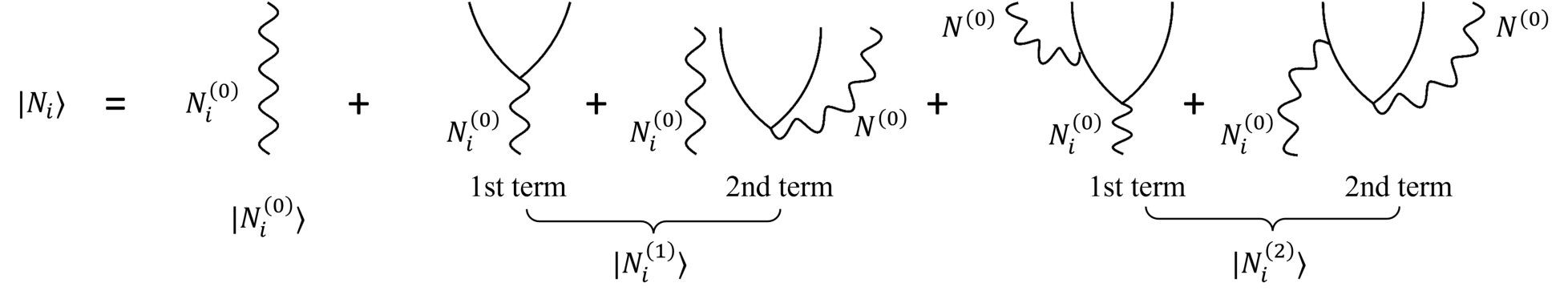}
    \caption{Perturbation expansion of the wave function $|N_i \rangle$.
        The wavy and solid lines represent the phonon and quasiparticle, respectively.
        Quasiparticle states are not explicitly labeled due to the possible different contractions among them, 
        as discussed in Appendix \ref{Apdx_qpvc_me}.}
    \label{fig:wf_corr}
\end{figure*}

We start from the Hamiltonian which describes a system that consists of interacting fermions and bosons \cite{Mottelson1976,Bortignon1977}
\begin{equation}
  H = H_{\rm F} + H_{\rm B} + V,
\end{equation}
where $H_{\rm F}$ is the Hamiltonian of single-quasiparticle states ($|a\rangle$) with energy $E_a$, 
and $H_{\rm B}$ is the Hamiltonian of QRPA phonons ($|N^{(0)}\rangle$) with energy $\Omega_N$, i.e., nuclear vibrations.
These two degrees of freedom are coupled through the interaction $V$.
Here $V$ includes both the particle-hole (ph) channel of the Skyrme interaction 
and the particle-particle (pp) channel of the pairing interaction.

For the $\gamma$ decay between two vibrational states $|N_i\rangle$ and $|N_f\rangle$,
we have to consider the interplay between nuclear collective motion and individual particles.
In other words, the collective mode dynamically modifies the nuclear mean field in which the particles move.
To consider this effect,
we work in a fixed quasiparticle basis and treat the interaction
$V$ as a perturbation.
The interaction dresses the QRPA phonons by 
admixing the configurations, such as 2qp and 2qp$\otimes$phonon,
to the QRPA states $|N^{(0)}\rangle$.
Accordingly, we expand the dressed states as
\begin{equation}
  |N\rangle = |N^{(0)}\rangle + |N^{(1)}\rangle + |N^{(2)}\rangle + \cdots,
\end{equation}
where $|N^{(k)} \rangle$ denotes the $k$-th order correction induced by $V$.
Up to second order, the perturbative expansion for the initial state $|N_i\rangle$ is
\begin{widetext}
\begin{subequations}
\begin{align}
|N_i^{(0)}\rangle =& |N_i^{(0)}\rangle  ;  \\
|N_i^{(1)}\rangle =& \sum_{ab} |ab\rangle \frac{\langle ab|V|N_i^{(0)}\rangle}{\Omega_{N_i} - (E_{a} + E_{b})}  \nonumber \\
                   &  + \sum_{abN} |abN^{(0)} N_i^{(0)}\rangle \frac{\langle abN^{(0)} N_i^{(0)}|V|N_i^{(0)}\rangle}
                         {\Omega_{N_i} - (\Omega_{N_i}+\Omega_{N}+E_{a} + E_{b})}  ; \\
|N_i^{(2)}\rangle =& \sum_{ab}\sum_{a'b'N} |a'b'{N}^{(0)}\rangle \frac{\langle a'b'{N}^{(0)}|V|ab\rangle}
                           {\Omega_{N_i} - (E_{a'} + E_{b'} + \Omega_{N})} 
                     \frac{\langle ab|V|N_i^{(0)}\rangle}{\Omega_{N_i} - (E_{a} + E_{b})} \nonumber \\
                   & + \sum_{abN}\sum_{a'b'}   |a'b'{N}^{(0)}\rangle \frac{\langle a'b'|V|ab N_i^{(0)}\rangle}
                           {\Omega_{N_i} - (E_{a'} + E_{b'} + \Omega_{N})} 
                     \frac{\langle ab{N}^{(0)} N_i^{(0)}|V|N_i^{(0)}\rangle}
                          {\Omega_{N_i} - (\Omega_{N_i}+\Omega_{N}+E_{a} + E_{b})}  .
\end{align}
\label{eq:wf_PT}
\end{subequations}
\end{widetext}
The diagrams in Fig. \ref{fig:wf_corr} provide a graphical representation of the above expressions.
For the final state, the expansion is quite similar.
We note that terms of pure ``bubble'' (ring) type in $|N^{(2)}\rangle$ are not considered, 
because their resummation is precisely what generates the QRPA phonon  $| N^{(0)} \rangle$.
In addition, some terms in $|N^{(2)}\rangle$ do not contribute at the second order of 
the transition matrix element $\langle N_f| Q_{\lambda \mu} | N_i \rangle$,
owing to a mismatch in the total number of creation and annihilation operators 
when combining these terms with the one-body transition operator.

The electric multipole operator reads
\begin{equation}
Q_{\lambda\mu}
= \sum_{i=1}^{A}
e^{\textrm{eff}}_{i}
r_{i}^{\lambda} Y_{\lambda\mu}(\hat{\bm{r}}_i),
\label{eq:Q_lm}
\end{equation}
where the effective charge caused by the recoil of the nucleus
has been introduced \cite{Bohr1998I}. In the dipole case with $\lambda = 1$,
$e_{\rm n}^{\textrm{eff}} = -Z/A$ for the neutron 
and $e_{\rm p}^{\textrm{eff}}= N/A$ for the proton.

\subsection{$\gamma$-decay width in NFT}
\label{subsec:bare}

\begin{figure*}[tb]
\centering
\includegraphics[width=0.9\textwidth]{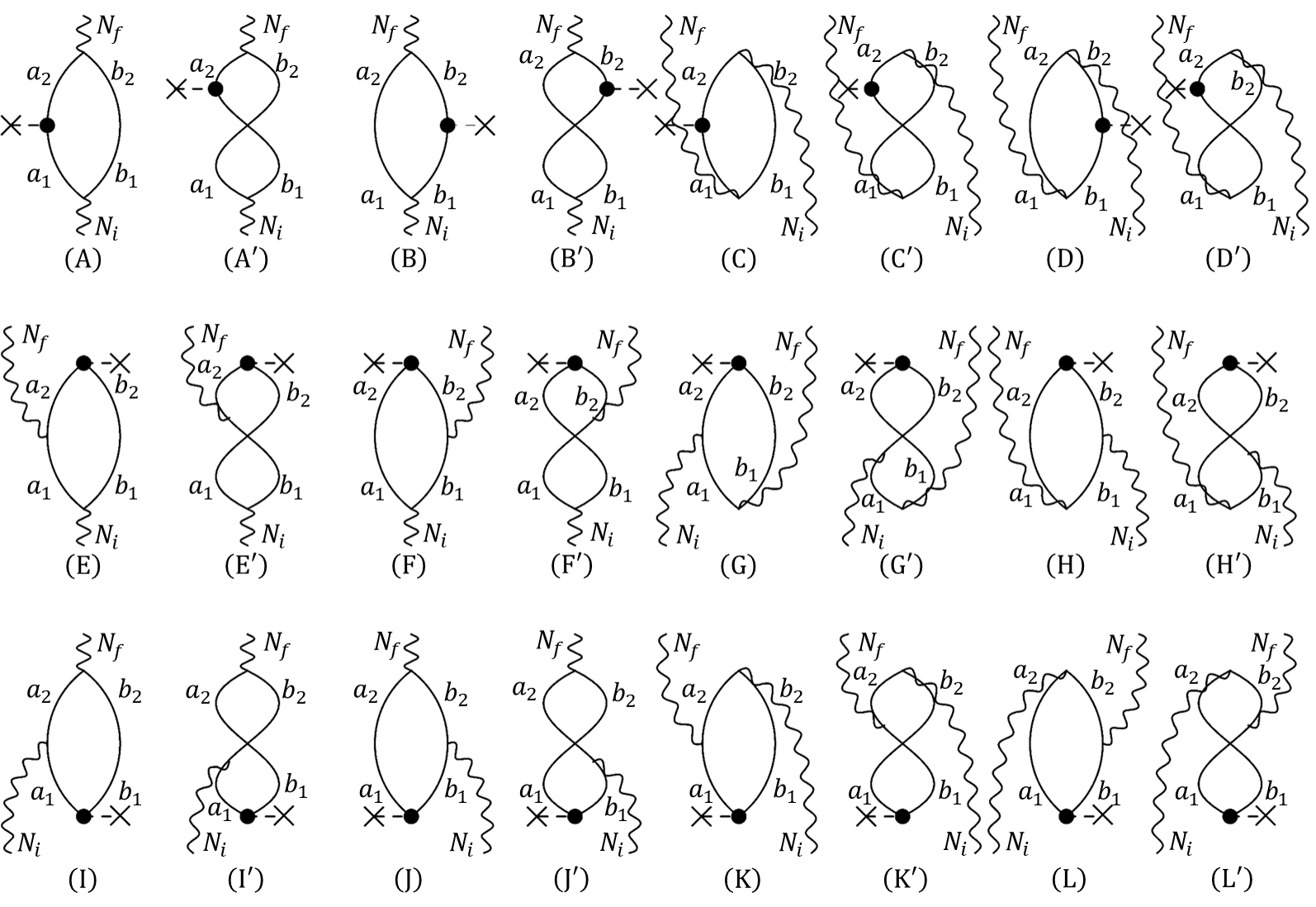}
\caption{The 24 second-order NFT diagrams for the $\gamma$-decay process between two vibrational states. 
The cross denotes the external operator $Q_{\lambda\mu}$. 
The shaded circle includes the contribution to $Q_{\lambda\mu}$ from nuclear polarization.}
\label{fig:fig_diagrams}
\end{figure*}

With the building blocks of the wave functions $|N\rangle$ [Eq. \eqref{eq:wf_PT}] 
and the one-body transition operator $Q_{\lambda\mu}$ [Eq. \eqref{eq:Q_lm}],
the transition matrix element between vibrational states can be written as
\begin{equation}
    \langle N_f | Q_{\lambda\mu} | N_i\rangle
  = \sum_{k=0}^{2} \sum_{k'=0}^{2} 
    \langle N_f^{(k')} | Q_{\lambda\mu} | N_i^{(k)} \rangle .
\end{equation}
In our model, we keep terms up to second order in the interaction $V$.
The non-vanishing matrix elements of a one-body operator $Q_{\lambda\mu}$ are of three types:
(i) single-quasiparticle transitions:
$\langle a| Q_{\lambda\mu} | b\rangle$,
(ii) quasiparticle-pair creation or annihilation, 
$\langle ab| Q_{\lambda\mu} |0\rangle$ or $\langle 0| Q_{\lambda\mu} |ab\rangle$,
(iii) phonon creation or annihilation, 
$\langle N^{(0)}| Q_{\lambda\mu} |0\rangle$ or $\langle 0| Q_{\lambda\mu} |N^{(0)}\rangle$.
Therefore, we can construct the Feynman diagrams contributing
to the $\gamma$ decay between two vibrational states in Fig. \ref{fig:fig_diagrams}.
Note that the phonon creation or annihilation of $Q_{\lambda\mu}$
will appear in the polarization effect, see Sec. \ref{subsec:polar}.

The reduced transition matrix element at the QRPA level is
$\langle N_f^{(1)} || Q_{\lambda\mu} || N_i^{(1)}\rangle$,
including contributions from diagrams A to D$'$,
\begin{widetext}
\begin{subequations}
	\begin{align}
   \langle N_f^{(1)} J_f || Q_{\lambda} || N_i^{(1)} J_i \rangle_{(m)}
=& 
   \sum_{a_1 b_1 a_2 b_2} 
   X^{N_f J_f \ast}_{a_2 b_2}
   X^{N_i J_i}_{a_1 b_1}  
   \langle (a_2 b_2)_{J_f} || Q_{\lambda} || (a_1 b_1)_{J_i} \rangle_{(m)} ,
   ~~ m = {\rm A, A', B, B'} \\
   \langle N_f^{(1)} J_f || Q_{\lambda} || N_i^{(1)} J_i \rangle_{(n)}
=& 
   (-)^{J_i + J_f + \lambda}
   \sum_{a_1 b_1 a_2 b_2} 
   Y^{N_i J_i}_{a_2 b_2}
   Y^{N_f J_f \ast}_{a_1 b_1}  
   \langle (a_2 b_2)_{J_i} || Q_{\lambda} || (a_1 b_1)_{J_f} \rangle_{(n)} ,
   ~~ n = {\rm C, C', D, D'}
\end{align}
\label{eq:dia_AD}
\end{subequations}
where $X_{ab}^{NJ}$ and $Y_{ab}^{NJ}$ are the forward and backward amplitudes 
in the QRPA model.
The detailed expressions of transition matrix elements 
$\langle (a_2 b_2)_{J_2} || Q_{\lambda} || (a_1 b_1)_{J_1} \rangle$
can be found in Appendix \ref{Apdx_qp_trans}.

The reduced transition matrix elements at QPVC level are
$\langle N_f^{(0)} || Q_{\lambda\mu} || N_i^{(2)}\rangle$ (diagrams E to H$'$) and
$\langle N_f^{(2)} || Q_{\lambda\mu} || N_i^{(0)}\rangle$ (diagrams I to L$'$),
\begin{subequations}
\begin{align}
		\langle N_f^{(0)} J_f || Q_{\lambda } || N_i^{(2)} J_i \rangle_{(p)}
		=& \hat{J}_i \hat{\lambda}^{-1}
		\sum_{a_1 b_1} \sum_{a_2 b_2} 
		X^{N_i J_i}_{a_1 b_1}
		\frac{A^{(3)J_i}_{a_2 b_2(\lambda)N_f,a_1 b_1}(p)}{[\Omega_{N_i} - (\Omega_{N_f} + E_{a_2} + E_{b_2})]+\mathrm{i}\eta_1}  
\langle 0 || Q_{\lambda} || (a_2 b_2)_{\lambda}  \rangle ,
~~p = {\rm E, E', F, F'}  \\
		\langle N_f^{(0)} J_f || Q_{\lambda} || N_i^{(2)} J_i \rangle_{(q)}
		=& \sum_{a_1 b_1} \sum_{a_2 b_2} 
		Y^{N_f J_f \ast}_{a_1 b_1} 
		\frac{A^{(2) \lambda}_{a_2 b_2, a_1 b_1 (J_f) N_i}(q)}{ [\Omega_{N_i} - (\Omega_{N_f} + E_{a_2} + E_{b_2})]+\mathrm{i}\eta_1 }  
\langle 0 || Q_{\lambda} || (a_2 b_2)_{\lambda}  \rangle ,
~~q = {\rm G, G', H, H'} \\
   \langle N_f^{(2)} J_f || Q_{\lambda} || N_i^{(0)} J_i \rangle_{(r)}
=& (-)^{J_i + J_f + \lambda} \hat{J}_f \hat{\lambda}^{-1}  
   \sum_{a_1 b_1} \sum_{a_2 b_2} 
   X^{N_f J_f \ast}_{a_2 b_2}
   \frac{A^{(2) J_f}_{a_2 b_2, a_1 b_1(\lambda)N_i}(r)}{ [\Omega_{N_f} - (\Omega_{N_i} + E_{a_1} + E_{b_1})]-\mathrm{i}\eta_1 }  
\langle  (a_1 b_1)_\lambda || Q_{\lambda} ||0 \rangle ,
~~r= {\rm I, I', J, J'}  \\
   \langle N_f^{(2)} J_f || Q_{\lambda} || N_i^{(0)} J_i \rangle_{(s)}
=&  (-)^{J_i + J_f + \lambda}
   \sum_{a_1 b_1} \sum_{a_2 b_2} 
   Y^{N_i J_i}_{a_2 b_2} 
   \frac{A^{(3) \lambda}_{a_2 b_2(J_i) N_f, a_1 b_1} (s)}{[\Omega_{N_f} - (\Omega_{N_i} + E_{a_1} + E_{b_1})]-\mathrm{i}\eta_1}  
\langle  (a_1 b_1)_\lambda || Q_{\lambda} ||0 \rangle ,
~~s = {\rm K, K', L, L'} 
\end{align}
\label{eq:dia_EL}
\end{subequations}
\end{widetext}
where $\hat{J} = \sqrt{2J+1}$.
The imaginary part $\eta_1$ in the energy denominator takes into account the coupling 
to more complicated configurations not included in our model space.
The detailed expressions for the quasiparticle transition matrix elements, 
as well as the QPVC matrix elements $A^{(2)}$ and $A^{(3)}$ can be found in Appendix \ref{Apdx_qp_trans} and \ref{Apdx_qpvc_me}.
Diagrams E to H$'$ correspond to the process of initial phonon $|N_i\rangle$ scattering into $|a_2 b_2 N_f \rangle$.
Therefore, they will contribute significantly if the initial GDR state 
includes configurations with 2qp coupled with the final $2_{1}^{+}$ state in its wave function.
The polarization effect will be presented in Sec. \ref{subsec:polar}.

By considering the 24 diagrams sketched in Fig. \ref{fig:fig_diagrams},
the transition strength can be calculated as
\begin{equation}
  B_{\gamma}^{fi}
= \frac{1}{2J_i +1} |\sum_{n = A, \dots, L'} \langle N_f|| Q_{\lambda} || N_i \rangle_{(n)} |^2.
\end{equation}
The corresponding $\gamma$-decay width $\Gamma_{\gamma}$ is 
\begin{equation}
	\Gamma_{\gamma}
	= \frac{1}{4\pi \epsilon_0} \frac{8\pi(\lambda+1)}{\lambda\left[(2\lambda+1)!!\right]^2}
	\left(\frac{\Delta E}{\hbar c}\right)^{2\lambda+1}
	B_{\gamma}^{fi},
	\label{Eq_width}
\end{equation}
where $\Delta E= \Omega_{N_i} - \Omega_{N_f}$ represents the transition energy. 

\subsection{Polarization effect}
\label{subsec:polar}

The large transition moment of $Q_{\lambda\mu}$ with $\lambda=1$ will induce nuclear vibrations \cite{Bohr1998II}. 
As a result, the initial and final states, $N_i$ and $N_f$ are clothed in a cloud of quanta.
This, in turn, gives rise to important modifications to the transition moment.
We deal with this kind of polarization effect using a factorization approach, as sketched in Fig. \ref{fig:fig_pol}.
The matrix element of $Q_{\lambda\mu}$ between the clothed phonons is
\begin{widetext}
\begin{equation}
\langle N_f | Q_{\lambda\mu}^{\rm pol} | N_i \rangle
\equiv
\langle \tilde{N}_f | Q_{\lambda\mu} | \tilde{N}_i \rangle
=  \langle N_f | Q_{\lambda\mu} | N_i \rangle
 + \sum_{\nu}
   \frac{ \langle 0 | Q_{\lambda\mu} | \nu \rangle 
          \langle N_f \nu | V | N_i \rangle  }
        { \Omega_{N_i} - (\Omega_{N_f} + \Omega_{\nu}) }
 + \frac{ \langle N_f | V | N_i \nu \rangle 
          \langle \nu | Q_{\lambda\mu} | 0 \rangle  }
        { \Omega_{N_f} - (\Omega_{N_i} + \Omega_{\nu}) }.
\label{eq:pol_trans_pho}
\end{equation}
Second, we further consider the interaction between phonon and individual quasiparticles,
so that we insert the expansion of $N_i$ and $N_f$ in Eq. \eqref{eq:wf_PT} into Eq. \eqref{eq:pol_trans_pho}.
We take diagrams A, E, and I as examples, 
because they are the diagrams that typically come from
$\langle N_f^{(1)} || Q_{\lambda\mu} || N_i^{(1)}\rangle$,
$\langle N_f^{(0)} || Q_{\lambda\mu} || N_i^{(2)}\rangle$,
and $\langle N_f^{(2)} || Q_{\lambda\mu} || N_i^{(0)}\rangle$,
respectively.
After considering the polarization effect they become 
\begin{figure*}[htbp]
\centering
\includegraphics[width=0.75\textwidth]{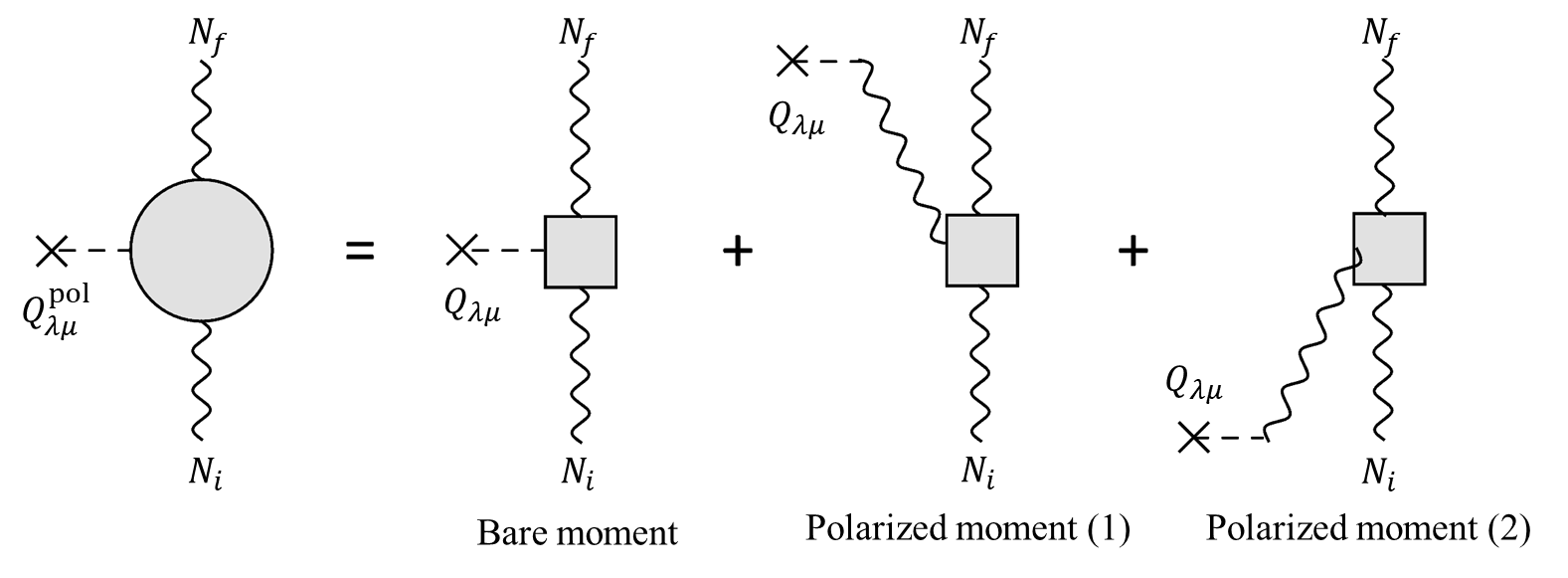}
\caption{
Diagrammatic representation of the polarization effect. 
The shaded square denotes the perturbative process shown in Fig.~\ref{fig:fig_diagrams}.
}
\label{fig:fig_pol}
\end{figure*}

\begin{figure*}[tbph]
    \centering
    \includegraphics[width=0.85\textwidth]{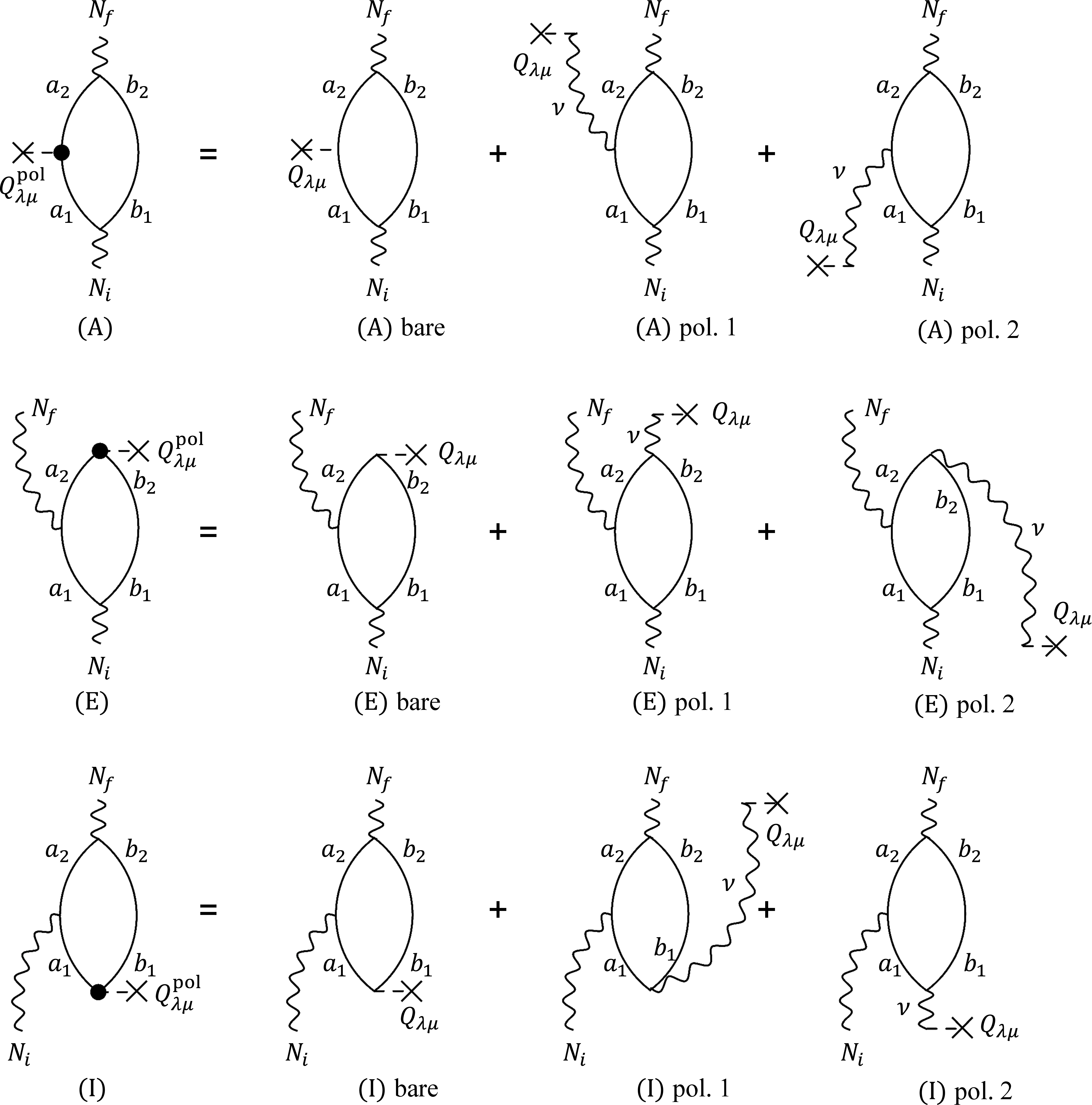}
    \caption{The polarization effect in diagram A, E, and I.}
    \label{fig:fig_polAEI}
\end{figure*}

\begin{equation}
\begin{aligned}
  \langle N_f^{(1)} || Q_{\lambda}^{\rm pol} || N_i^{(1)} \rangle_{(\rm A)}
=& \langle N_f^{(1)} || Q_{\lambda} || N_i^{(1)} \rangle_{(\rm A)} 
 \\ & 
 + \hat{\lambda}^{-1} \sum_{\nu} 
      (-)^{J_i + J_f + \lambda} \hat{J_i}
        \frac{\langle 0 || Q_{\lambda} || \nu \rangle
            X_{a_2 b_2}^{N_f J_f \ast}
            [A^{(3) J_i}_{a_2 b_2 (J_f) \nu, a_1 b_1}(1)]
            X_{a_1 b_1}^{N_i J_i} }
        {[\Omega_{N_i} - (\Omega_{N_f} + \Omega_{\nu}) + \mathrm{i} \eta_2] } 
 \\ & \qquad
       + \hat{J}_f \frac{
           X_{a_2 b_2}^{N_f J_f  \ast}
           [A^{(2) J_f}_{a_2 b_2, a_1 b_1 (J_i) \nu}(1)]
           X_{a_1 b_1}^{N_i J_i} 
           \langle \nu || Q_{\lambda} || 0 \rangle }
        {[\Omega_{N_f} - (\Omega_{N_i} + \Omega_{\nu}) - \mathrm{i} \eta_2] } ,
    \end{aligned}
\end{equation}

\begin{equation}
\begin{aligned}
  \langle N_f^{(0)} || Q_{\lambda}^{\rm pol} || N_i^{(2)} \rangle_{(\rm E)}
=& \langle N_f^{(0)} || Q_{\lambda} || N_i^{(2)} \rangle_{(\rm E)} \\
 &+ \hat{\lambda}^{-1} \sum_{\nu}
 \frac{ \langle 0 || Q_{\lambda} || \nu \rangle
        \langle \nu || V || a_2 b_2\rangle
        [A^{(3) J_i}_{a_2 b_2 (\lambda) N_f, a_1 b_1} (1)]
        X_{a_1 b_1}^{N_i J_i} }
       {[\Omega_{N_i} - (\Omega_{N_f} + \Omega_{\nu}) + \mathrm{i} \eta_2]
        [\Omega_{N_i} - (E_{a_2} + E_{b_2} + \Omega_{N_f})]} \\
 &\qquad
 +\frac{\langle 0 || V || a_2 b_2 \nu \rangle
        [A^{(3) J_i}_{a_2 b_2 (\lambda) N_f, a_1 b_1 }(1)]
        X_{a_1 b_1}^{N_i J_i} 
        \langle \nu || Q_{\lambda} || 0 \rangle } 
       {[\Omega_{N_f} - (\Omega_{N_i} + \Omega_{\nu}) - \mathrm{i} \eta_2]
        [\Omega_{N_i} - (E_{a_2} + E_{b_2} + \Omega_{N_f})]} ,
\end{aligned}
\end{equation}

\begin{equation}
    \begin{aligned}
        \langle N_f^{(2)} || Q_{\lambda}^{\rm pol} || N_i^{(0)} \rangle_{(\rm I)}
        =& \langle N_f^{(2)} || Q_{\lambda} || N_i^{(0)} \rangle_{(\rm I)} \\
        &+ \hat{\lambda}^{-1} \sum_{\nu}
        \frac{ \langle 0 || Q_{\lambda} || \nu \rangle
            X_{a_2 b_2}^{N_f J_f \ast}
            [A^{(2) J_f}_{a_2 b_2, a_1 b_1 (\lambda) N_i} (1)]
            \langle a_1 b_1 \nu || V || 0\rangle }
        {[\Omega_{N_i} - (\Omega_{N_f} + \Omega_{\nu}) + \mathrm{i} \eta_2]
         [\Omega_{N_f} - (E_{a_1} + E_{b_1} + \Omega_{N_i})] } \\
        &\qquad
        +\frac{ X_{a_2 b_2}^{N_f J_f \ast}
         [A^{(2) J_f}_{a_2 b_2, a_1 b_1 (\lambda) N_i} (1)]
         \langle a_1 b_1 || V || \nu\rangle 
         \langle \nu || Q_{\lambda} || 0 \rangle }
        {[\Omega_{N_f} - (\Omega_{N_i} + \Omega_{\nu}) - \mathrm{i} \eta_2]
         [\Omega_{N_f} - (E_{a_1} + E_{b_1} + \Omega_{N_i})] } .
    \end{aligned}
\end{equation}
Figure \ref{fig:fig_polAEI} represents the above expressions in a diagrammatic way.
The imaginary part $\eta_2$ accounts for damping of the phonon $\nu$.
Expressions of the QRPA vertices $\langle \nu || V || a b \rangle$ and $\langle ab,\nu || V || 0 \rangle$
are presented in Appendix \ref{Apdx_vertex}.
Polarization corrections in the remaining diagrams can be obtained analogously.

By comparing the above equations with Eqs. \eqref{eq:dia_AD} and \eqref{eq:dia_EL}, 
the polarization effect can be included in the transition matrix elements of quasiparticles.
For diagrams A to D$'$, 
the polarized transition matrix elements are 
\begin{subequations}
  \begin{align}
   \langle (a_2 b_2)_{J_2} || Q_{\lambda}^{\rm pol} || (a_1 b_1)_{J_1} \rangle_{(m)}
=& \langle (a_2 b_2)_{J_2} || Q_{\lambda} || (a_1 b_1)_{J_1} \rangle_{(m)}   \nonumber \\ &
 + \hat{\lambda}^{-1} \sum_{\nu}
 (-)^{J_1 + J_2 + \lambda} \hat{J_1}
 \frac{ \langle 0 || Q_{\lambda} || \nu \rangle [A^{(3) J_1}_{a_2 b_2 (J_2) \nu, a_1 b_1}(n)]  }
 { \Omega_{N_i} - (\Omega_{N_f} + \Omega_{\nu}) + \mathrm{i}\eta_2 } 
 + \hat{J_2} 
 \frac{ [A^{(2) J_2}_{a_2 b_2, a_1 b_1 (J_1) \nu}(n)] \langle \nu || Q_{\lambda} || 0 \rangle }
 { \Omega_{N_f} - (\Omega_{N_i} + \Omega_{\nu}) - \mathrm{i}\eta_2}  ,
\end{align}
\end{subequations}
where $m={\rm A},{\rm A}',{\rm B},{\rm B}'$ and $m={\rm C}, {\rm C}', {\rm D}, {\rm D}'$ correspond to $n=1,2,3,4$.

For diagrams E to H$'$, and I to L$'$, 
the polarized transition matrix elements are respectively
\begin{subequations}
    \begin{align}
        \langle 0|| Q_{\lambda}^{\rm pol} || (a_2 b_2)_{\lambda} \rangle
        =& \langle 0|| Q_{\lambda} || (a_2 b_2)_{\lambda} \rangle 
        + \hat{\lambda}^{-1} \sum_{\nu} 
        \frac{ \langle 0|| Q_{\lambda} ||\nu \rangle \langle \nu || V || a_2 b_2 \rangle  }
        { \Omega_{N_i} - (\Omega_{N_f} + \Omega_{\nu}) + \mathrm{i}\eta_2}
        +  
        \frac{ \langle 0 || V || a_2 b_2 \nu \rangle \langle \nu || Q_{\lambda} ||0 \rangle}
        { \Omega_{N_f} - (\Omega_{N_i} + \Omega_{\nu}) - \mathrm{i}\eta_2}  , \\
        \langle (a_1 b_1)_{\lambda}|| Q_{\lambda}^{\rm pol} || 0 \rangle
        =& \langle (a_1 b_1)_{\lambda}|| Q_{\lambda} || 0 \rangle 
        + \hat{\lambda}^{-1} \sum_{\nu}  
        \frac{ \langle 0|| Q_{\lambda} ||\nu \rangle \langle a_1 b_1 \nu || V || 0 \rangle }
        { \Omega_{N_i} - (\Omega_{N_f} + \Omega_{\nu}) + \mathrm{i}\eta_2}
        +
        \frac{ \langle a_1 b_1 || V || \nu \rangle \langle \nu\lambda|| Q_{\lambda} || 0 \rangle }
        { \Omega_{N_f} - (\Omega_{N_i} + \Omega_{\nu}) - \mathrm{i}\eta_2} .
    \end{align}
\end{subequations}
\end{widetext}

\section{Numerical details}
\label{secNume}

The phonon states are calculated with the self-consistent QRPA model.
The box size for calculating the single-particle levels is 20 fm.
A smooth cut-off of 60 MeV is applied to the equivalent Hartree-Fock energy of the quasiparticle states, 
having a Fermi profile of diffuseness 0.1 MeV.
Volume pairing is adopted consistently in 
HFB, in the QRPA calculation, and in all the vertices of the diagrams.
{
The pairing strengths are adjusted 
to reproduce the mean pairing gap estimated from 
the standard three-point mass-difference formula \cite{Bender2000}.
For the proton pairing channel in $^{140}$Ce,
the empirical mean pairing gap $\langle\Delta_{\rm p}\rangle$ is 1.46 MeV. 
Since $^{140}$Ce has the neutron magic number $N=82$, 
the neutron pairing strength is taken to be the same as
the proton pairing strength in the residual interaction.}
The values for the volume pairing strengths are respectively 
$-206.1$ MeV$\cdot$fm$^{3}$, $-198.1$ MeV$\cdot$fm$^{3}$,
$-186.7$ MeV$\cdot$fm$^{3}$, and $-182.9$ MeV$\cdot$fm$^{3}$
for SIII \cite{Beiner1975SII}, SGII \cite{VanGiai1981SGII},
SkM$^{\ast}$ \cite{Bartel1982SkMst}, and LNS \cite{Cao2006LNS}.
{The model space is sufficiently large to exhaust essentially the whole isovector energy-weighted sum rule, 
    with values of 99.9\%-100.1\% for these four Skyrme interactions.}
The GDR states are chosen from the QRPA states within the energy range of 10-20 MeV, 
requiring a fraction of isovector (IV) non-energy-weighted sum rule (NEWSR) larger than 5\%. 
The selection criteria for the dipole modes that contribute to the polarization effect
are IV or isoscalar (IS) NEWSR fraction larger than 5\% and energy smaller than 30 MeV.
{
The NEWSR 5\% cutoff has been used in our earlier calculations of $\gamma$ decay \cite{Brenna2012,Lv2021},
and in QPVC studies of 
charge-exchange and non-charge-exchange nuclear excitations \cite{Niu2014,Niu2015,Niu2016}.}
{
Both IS and IV phonons are included in the polarization effect. 
We find that the polarization effect from IS phonons on the decay width is less than 1\%.
}

\begin{figure}[tb]
\centering
\includegraphics[width=0.45\textwidth]{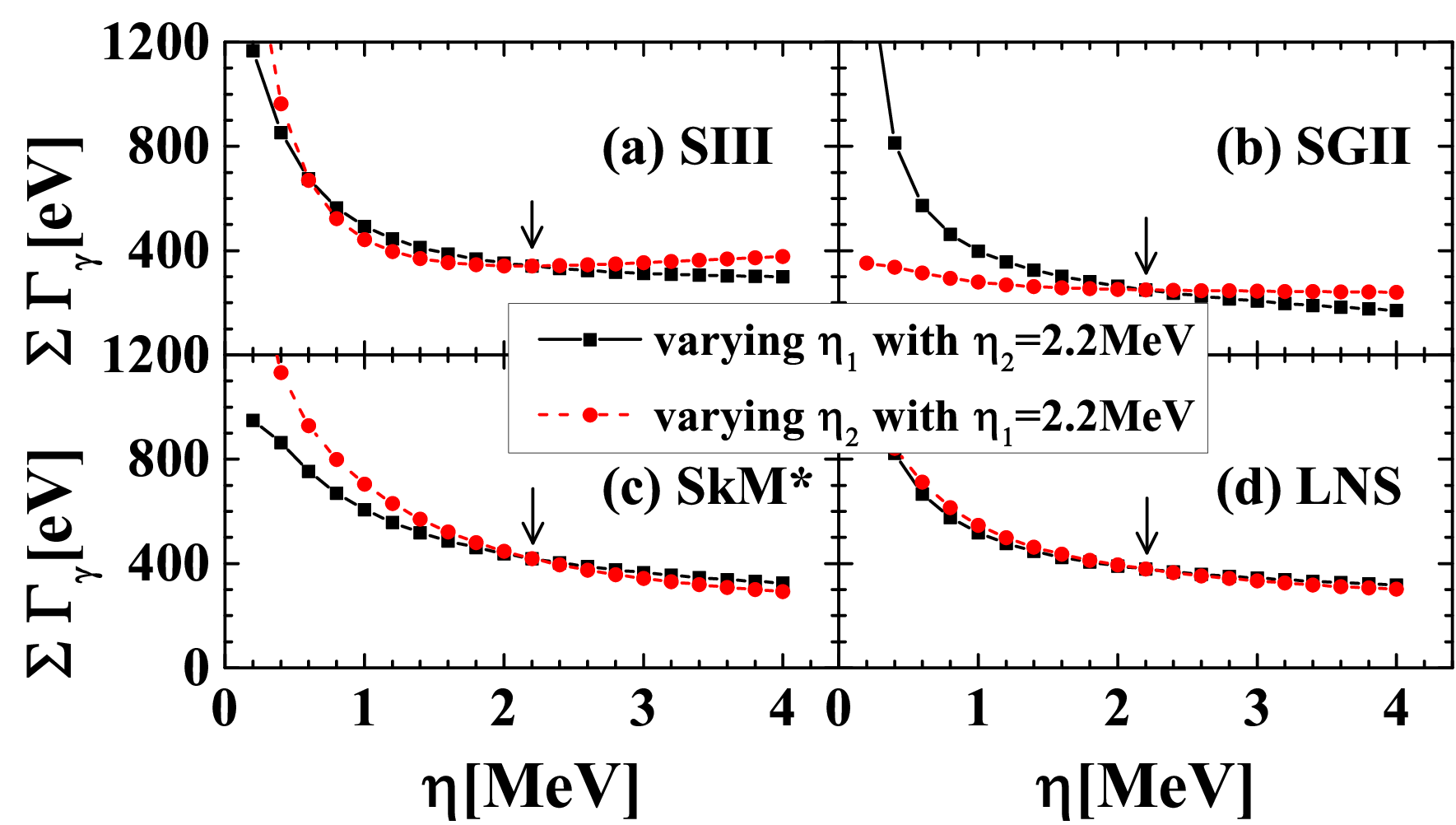}
\caption{Sensitivity of the results for the $\gamma$-decay width against variation of the imaginary parts $\eta_1$ and $\eta_2$
for SIII (a), SGII (b), SkM* (c), and LNS (d) functionals. 
$\sum \Gamma_{\gamma}$ denotes the sum of the $\gamma$-decay width to the $2_{1}^{+}$ state for the selected dipole modes in GDR region.
Black (red) line shows the result of varying $\eta_1$ ($\eta_2=2.2$ MeV) while keeping $\eta_1=2.2$ MeV ($\eta_2=2.2$ MeV).
}
\label{fig:fig_eta_all}
\end{figure}

{
We investigate the sensitivity of the results against variation of the two imaginary parts $\eta_1$ and $\eta_2$ 
by studying the dependence of $\sum \Gamma_\gamma$ on their values for the 4 Skyrme functionals.
Here $\sum \Gamma_\gamma$ is the summed $\gamma$-decay width 
to the $2_{1}^{+}$ state for the selected dipole modes within the GDR region.
As shown in Fig. \ref{fig:fig_eta_all}, 
the $\gamma$-decay width varies rapidly at small values of $\eta_1$ and $\eta_2$, 
whereas the dependence becomes substantially weaker around and above $\eta_1=\eta_2\simeq 2.2$ MeV.
The similar dependence of $\sum\Gamma_\gamma$ on $\eta_1$ and on $\eta_2$ is not accidental,
because both of them enter the denominators in a similar way and mimic the damping of complex configurations.
We then adopt $\eta_1=\eta_2=2.2$ MeV, motivated by the experimental GDR width in $^{140}$Ce, 
for which 2.2 MeV corresponds to one half of the measured
full width at half maximum~\cite{Lepretre1976}.
}

\section{Results and discussions}
\label{secResu}

\begin{table}[tb]
\centering
\caption{Experimental \cite{Peker1994} and theoretical excitation energies and $E2$ transition strengths of the $2_{1}^{+}$ state in $^{140}$Ce.
The energies of the GDR are listed in the last line, 
where the experimental value is obtained via standard Lorentzian fitting to the photoneutron cross section \cite{Lepretre1976}, 
while the theoretical ones are obtained by $m_1(E1) / m_0(E1)$.}
\begin{tabular}{cccccc}
\hline \hline
                      &  Expt.  & SIII  & SGII  &  SkM$^{\ast}$ &  LNS     \\ \hline
$E(2_1^{+})$[MeV]     & 1.60    & 2.50  & 2.10  &  1.80         & 2.07     \\ 
$B(E2)_{2_1^{+}}$[$10^{4}e^2{\rm fm}^2$]  
                      & 0.304   & 0.300 & 0.389 &  0.463        & 0.351    \\ 
$E({\rm GDR})$ [MeV]  & 15.03   & 16.30 & 15.20 &  15.47        & 15.73    \\
	\hline \hline
\end{tabular}
\label{tab:tab1}
\end{table}

\begin{figure}[tb]
\centering
\includegraphics[width=0.5\textwidth]{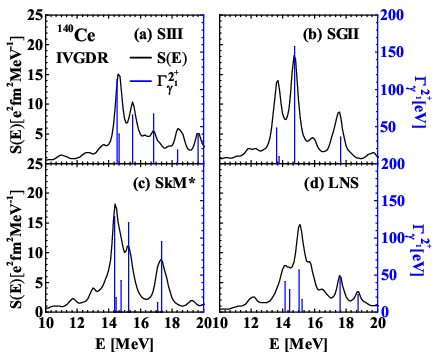}
\caption{$E1$ transition strength distributions (black lines) around the GDR region calculated by the QRPA model 
with SIII (a), SGII (b), SkM$^{\ast}$ (c), and LNS (d) interactions.
The width parameter $\Gamma$ of the Lorentzian function is set as 0.5 MeV to resolve the individual peaks clearly.
The partial $\gamma$-decay width to $2_1^+$ state $\Gamma_{\gamma}^{2_1^+}$ (blue square) of individual GDR states is also shown in the figure.
}
\label{fig:fig_gdr_sth}
\end{figure}

\begin{table*}[tb]
    \centering
    \caption{$\gamma$-decay widths and the branching ratios of the selected GDR states.
        The ratio $R$ for individual GDR state is calculated by $\Gamma_{\gamma}^{2_{1}^{+}} / \Gamma_{\gamma}^{\rm g.s.}$.
        The total ratio $R$ is calculated by $\sum \Gamma_{\gamma}^{2_{1}^{+}} / \sum \Gamma_{\gamma}^{\rm g.s.}$. }
    \begin{tabular}{cccccccccccccccc}
        \hline \hline
        SIII    &                                     &                                  &           &  SGII    &                                      &                                  &        &  SkM$^{\ast}$&                                 &                                  &           &  LNS     &                                      &                                  &              \\ \hline
        $E$[MeV]&  $\Gamma_{\gamma}^{2_{1}^{+}}$[eV]  & $\Gamma_{\gamma}^{\rm g.s.}$[eV] &  $R$[\%]  & $E$[MeV] &  $\Gamma_{\gamma}^{2_{1}^{+}}$[eV]   & $\Gamma_{\gamma}^{\rm g.s.}$[eV] & $R$[\%]&  $E$[MeV]&  $\Gamma_{\gamma}^{2_{1}^{+}}$[eV]  & $\Gamma_{\gamma}^{\rm g.s.}$[eV] &  $R$[\%]  & $E$[MeV] &  $\Gamma_{\gamma}^{2_{1}^{+}}$[eV]   & $\Gamma_{\gamma}^{\rm g.s.}$[eV] & $R$[\%]      \\
        14.53   &  112.9                              & 2952.9                           &  3.82     &  13.62   &   47.5                               & 4029.6                           & 1.18   &  14.38   &  128.3                              & 8209.0                           &  1.56     &  14.19   &   25.3                               & 4764.5                           & 0.53         \\
14.65   &   39.8                              & 9174.5                           &  0.43     &  13.77   &    9.5                               & 2624.9                           & 0.36   &  14.45   &   19.6                              & 3642.8                           &  0.54     &  15.03   &   60.4                               & 7376.2                           & 0.82         \\
15.50   &   65.4                              & 7491.7                           &  0.87     &  14.73   &  157.3                               &12744.4                           & 1.23   &  14.78   &   42.4                              & 4084.2                           &  1.04     &  15.15   &  137.2                               & 3121.8                           & 4.40         \\
16.84   &   66.4                              & 4400.8                           &  1.51     &  17.65   &   35.5                               & 6937.5                           & 0.51   &  15.26   &  120.8                              & 7547.8                           &  1.60     &  15.76   &   65.8                               & 3414.8                           & 1.93         \\ 
18.37   &   18.0                              & 6532.9                           &  0.28     &          &                                      &                                  &        &  17.10   &   12.7                              & 5141.9                           &  0.25     &  17.64   &   57.6                               & 8423.6                           & 0.68         \\
19.65   &   38.6                              & 8592.5                           &  0.45     &          &                                      &                                  &        &  17.34   &   95.2                              & 6887.4                           &  1.38     &  18.76   &   31.8                               & 5426.3                           & 0.59          \\ 
Total   &  341.0                              &39145.2                           &  0.87     &  Total   &  249.9                               &26336.5                           & 0.95   &  Total   &  419.0                              &35513.1                           &  1.18     &  Total   &  378.3                               &32527.2                           & 1.16          \\ 
        \hline
        \hline
    \end{tabular}
    \label{tab:tab2}
\end{table*}

The transition amplitude $\langle N_f J_f || Q_{\lambda} || N_i J_i \rangle$ involves both the initial and final states.
Therefore, we first assess the quality of the theoretical description of the $2_{1}^{+}$ state and GDR.
The experimental excitation energies and $B(E2)$ value, 
along with the QRPA results calculated with 
SIII \cite{Beiner1975SII}, SGII \cite{VanGiai1981SGII},
SkM$^{\ast}$ \cite{Bartel1982SkMst}, and LNS \cite{Cao2006LNS} Skyrme functionals, are listed in Table \ref{tab:tab1}.
The theoretical GDR energy is calculated by the ratio of 
energy-weighted sum rule to the non-energy-weighted sum rule $m_{1}(E1)/m_{0}(E1)$,
where the states in 8 to 28 MeV are included to match the Lorentzian fitting range of the experimental photoneutron cross section.
The $E1$ strength distributions 
of the 4 Skyrme functionals are depicted in Fig. \ref{fig:fig_gdr_sth}.
For the SIII functional, the energies for both $2_{1}^{+}$ and GDR are slightly higher than the experimental ones,
while the $B(E2)$ value for $2_{1}^{+}$ is reproduced very well.
SGII, SkM$^{\ast}$, and LNS functionals provide a better description of the excitation energies than SIII,
although they overestimate the $B(E2)$ value of the $2_{1}^{+}$ state.
Overall, these 4 Skyrme functionals provide a reasonable description of both the $2_{1}^{+}$ state 
and the GDR, and can thus be employed in the subsequent calculations of $\gamma$-decay.

Table \ref{tab:tab2} presents the $\gamma$-decay properties from the selected GDR states
to the first $2_1^+$ state in $^{140}$Ce calculated by SIII, SGII, SkM*, and LNS functionals. 
For each functional, the excitation energy $E$ of the single GDR state,
the corresponding partial $\gamma$-decay width to the $2_1^+$ state, $\Gamma_{\gamma}^{2_1^+}$, 
and the branching ratio are listed.
The total decay width $\sum \Gamma_{\gamma}^{2_{1}^{+}}$ varies from 249.9 eV (SGII) to 419.0 eV (SkM$^\ast$), 
while the total branching ratio ranges from 0.87\% to 1.18\%. 
{
However, according to the Weisskopf estimate \cite{Bohr1998I}, 
a typical single-particle dipole transition with an energy of 13.4 MeV,
corresponding to the experimental energy difference between the GDR and the
$2_1^+$ state in $^{140}$Ce, would produce a $\gamma$-decay width of about
$4400$ eV. 
The microscopic calculation presented here yields 
total decay widths in the range of 250-420 eV,
corresponding to a hindrance factor of about 10-20.
This strong reduction with respect to the Weisskopf estimate 
is an important feature of the present model,
leading to theoretical branching ratios 
in semi-quantitative agreement with the {preliminary experimental data} \cite{Kleemann2024}.
The hindrance arises from several competing effects
in the microscopic transition amplitude
\cite{Bortignon1984,Brenna2012,Lv2021}.
First, the angular momentum recoupling coefficients 
in Eqs. \eqref{eq:dia_AD} and \eqref{eq:dia_EL} 
reduce the contribution of individual quasiparticle configurations.
Second, particle-like and hole-like contributions partially cancel each other.
States with large occupation probabilities retain predominantly hole-like character, 
whereas those with small occupation probabilities are mainly particle-like, 
and the corresponding terms enter the transition amplitude with opposite signs.
Compared with the particle-hole cancellation in closed-shell nuclei, 
pairing correlations in the present open-shell system smear the distinction 
between particle-like and hole-like configurations 
and thereby weaken the associated hindrance.
In contrast, the interference between the proton and neutron
contributions enhances the decay width. 
In the present case, the isovector GDR is connected to the predominantly isoscalar
$2_1^+$ state by the isovector $E1$ operator, 
which gives opposite weights to the proton and neutron terms. 
Combined with the opposite phases of the proton and neutron transition
densities in the IVGDR, 
this results in predominantly constructive interference in the
$\mathrm{GDR}\rightarrow 2_1^+$ transition amplitude.
Finally, the polarization of the $E1$ transition moment through coupling to
dipole phonons further modifies the decay width. Depending on the relative
magnitudes of the transition energy $\Delta E$ and the GDR energy, the
polarization effect may either suppress or enhance the decay width. 
This effect will be discussed in detail later.
}
The state-by-state values of $\Gamma_{\gamma}^{2_1^+}$ are shown as blue squares in Fig.~\ref{fig:fig_gdr_sth}.
Overall, $\Gamma_{\gamma}^{2_1^+}$ shows a similar energy dependence to the $E1$ strength, 
exhibiting enhanced values around the main IVGDR peak.

\begin{figure}[b]
\centering
\includegraphics[width=0.45\textwidth]{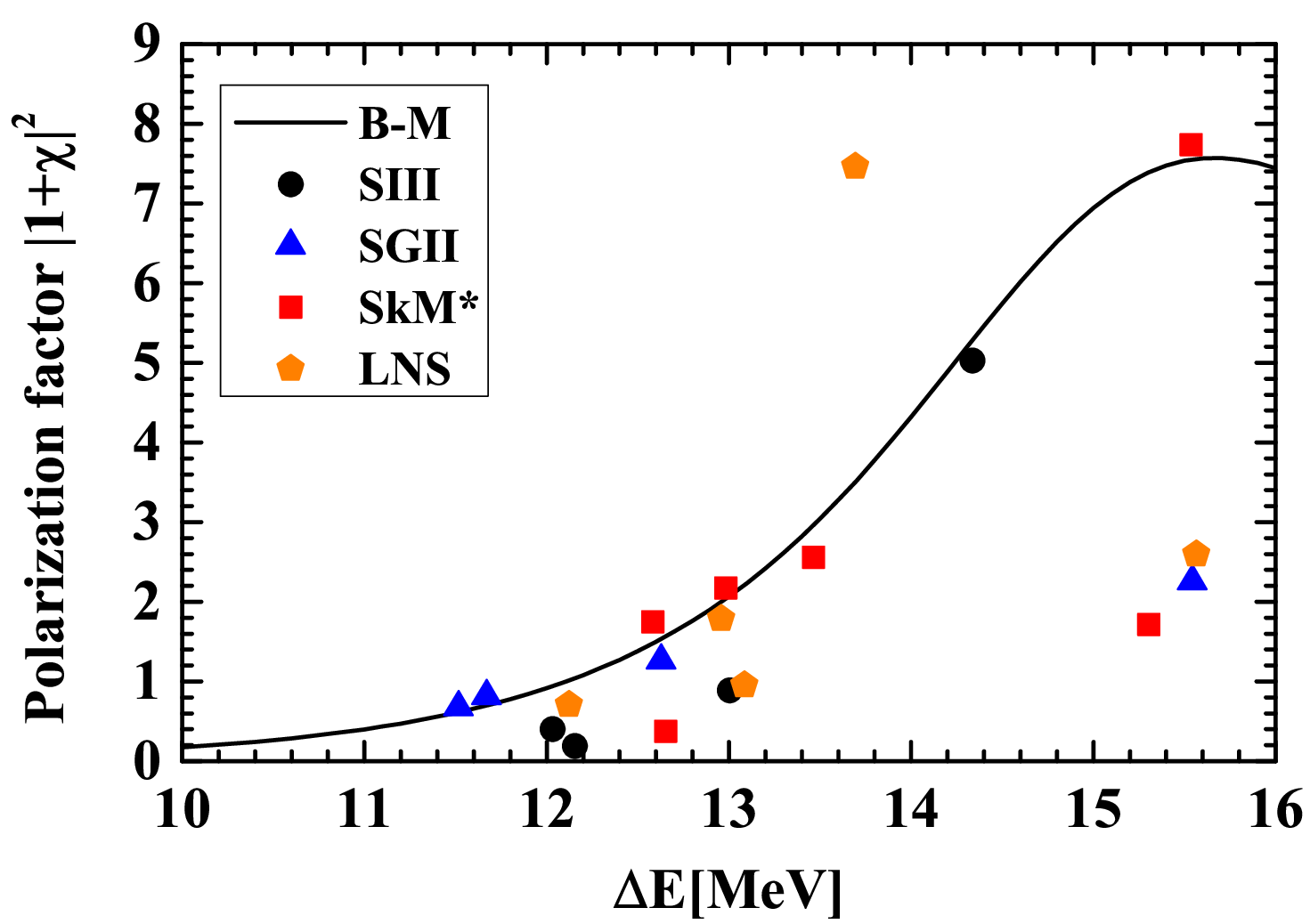}
\caption{
Polarization factor as a function of the transition energy $\Delta E = \Delta\Omega_{if}$.
The solid line is Bohr-Mottelson formula with experimental GDR energy and width.
The points are calculated by our model for the dipole states lying in the GDR region with different Skyrme functionals.
}
\label{fig:fig_pol_fac}
\end{figure}

In Fig. \ref{fig:fig_pol_fac}, we present the polarization factor, $|1+\chi|^2$, 
as a function of the transition energy $\Delta E= \Delta \Omega_{if}$, in the GDR region in $^{140}$Ce.
In our model, the polarization factor $|1+\chi|^2$  is calculated as  
the ratio of the $\gamma$-decay width with and without the polarization correction.
Macroscopically, the dipole polarizability $\chi$ can be calculated by the Bohr-Mottelson (B-M) formula
\cite{Bohr1998II,Bortignon1984}
\begin{equation}
\chi = \frac{-0.76 E_{\rm GDR}^2}
{ (E_{\rm GDR} + \Delta E + \mathrm{i} \frac{\Gamma_{\rm GDR}}{2})
(E_{\rm GDR} - \Delta E - \mathrm{i} \frac{\Gamma_{\rm GDR}}{2})  },
\end{equation}
where the experimental values $E_{\rm GDR} = 15.03$ MeV and $\Gamma_{\rm GDR}=4.4$ MeV are used.
The polarization factors calculated from the microscopic model 
generally follow the trend predicted by the macroscopic B-M formula. 
As the transition energy $\Delta E$ is well below the GDR energy,
the factor is smaller than 1, indicating that the polarization suppresses the $\gamma$-decay width. 
When $\Delta E$ increases and approaches the resonance region, the factor exceeds 1, 
implying that the polarization now enhances the decay width.
A maximum enhancement can be reached when $\Delta E$ is comparable to the GDR energy, 
where the virtual excitation of the collective mode is most efficient \cite{Bohr1998II}.
We note that there are some deviations from the B-M curve for several individual dipole states, 
arising from the more detailed treatment in our microscopic approach. 
Namely, the B-M formula essentially models the polarization effect mediated by a single collective GDR phonon,
while our calculation includes the contributions from several fragmented dipole phonon states.

\section{Summary}
\label{secSum}

In summary, we have developed a fully self-consistent Skyrme QPVC model
to calculate the $\gamma$ decay between vibrational states in superfluid nuclei.
We treat the initial and final states as QRPA phonons.
Then, within the framework of nuclear field theory, 
we consider the perturbation expansion of the phonon wave functions,
induced by the interaction between quasiparticles and phonons,
to ensure a comprehensive treatment of the many-body dynamics.
Moreover, we consider the important polarization effect. 
The self-consistency of our model lies in 
employing the same Skyrme force and pairing interaction, without further adjustable parameters,
both in the ground state, in the construction of the phonons, and in the interaction vertices. 

Motivated by recent experiments at the HI$\gamma$S~\cite{Kleemann2024},
we apply our model to the $\gamma$-decay from the GDR to the $2_{1}^{+}$ state in $^{140}$Ce,
with SIII, SGII, SkM$^{\ast}$ and LNS functionals.
{The excitation energies of the GDR and the $2_1^+$ state, 
as well as the $B(E2;{\rm g.s.} \rightarrow 2_1^+)$ value, are reasonably reproduced.}
For the decay process, we calculate the partial and total $\gamma$-decay widths 
of individual dipole modes within the GDR region. 
The total width ranges from approximately 250 to 420 eV, 
and the corresponding branching ratio varies between 0.87\% and 1.18\%, depending on the functional,
{which are in semiquantitative agreement with the preliminary experimental results.}
Furthermore, the polarization factor extracted from our microscopic model is compared with the Bohr-Mottelson formula.
Both approaches yield the same increasing trend with the transition energy.

\section*{Acknowledgements}

W.-L. L. acknowledges helpful discussions with Zheng-Zheng Li and Yi-Wei Hao.
Y.-F. N. and G. C.  acknowledge helpful discussions with Norbert Pietralla, Johann Isaak and the group at TU Darmstadt. 
This work was supported by 
the ``Young Scientist Scheme'' of National Key Research and Development (R\&D) Program under grant No. 2021YFA1601500,
the National Natural Science Foundation of China under grant Nos. 12405135, 12447168, 12075104, 12447106, 
the Science and Technology Innovation Leading Talent Project of Gansu Province (25RCKA025), 
the Lingchuang Research Project of China National Nuclear Corporation (CNNC-LCKY-2024-082), 
the Fundamental Research Funds for the Central Universities (lzujbky-2023-stlt01).

\begin{widetext}
\appendix
\section{Quasiparticle transition matrix elements}
\label{Apdx_qp_trans}

In the quasiparticle picture, the one-body operator can be written as
\begin{equation}
Q_{\lambda\mu}
= \sum_{ab} \sum_{m_a m_b} \hat{\lambda}^{-1}
  C_{j_a m_a j_b m_b}^{\lambda\mu}
  \langle j_a ||Q_{\lambda}|| j_b \rangle (
  -  u_{a} v_{b} \alpha^{\dag}_{a} \alpha^{\dag}_{b}
  +  u_{a} u_{b} \alpha^{\dag}_{a} \alpha_{\bar{b}}
  -  v_{a} v_{b} \alpha_{\bar{a}} \alpha^{\dag}_{b}
  +  v_{a} u_{b} \alpha_{\bar{a}} \alpha_{\bar{b}} ),
\end{equation}
where we use the special Bogoliubov transformation from particle to quasiparticle.

The two terms $\alpha^{\dag}_{a} \alpha_{\bar{b}}$ and $\alpha_{\bar{a}} \alpha^{\dag}_{b}$ in $Q_{\lambda\mu}$
result in 8 different contractions in the transition matrix element
$\langle (a_2 b_2)_{J_2}|| Q_{\lambda} || (a_1 b_1)_{J_1} \rangle$ \cite{Suhonen2007},
\begin{equation}
		\langle (a_2 b_2)_{J_2}|| Q_{\lambda} || (a_1 b_1)_{J_1} \rangle_{(1)}
		= \mathcal{N}_{a_1 b_1}^{J_1} \mathcal{N}_{a_2 b_2}^{J_2}
		\hat{J_1} \hat{J}_2
		\delta_{b_1 b_2}
		u_{a_2} u_{a_1} (-)^{j_{a_2} + j_{b_1} + J_1 + \lambda}
		\langle j_{a_2} ||Q_{\lambda}|| j_{a_1} \rangle
		\left\{
		\begin{array}{ccc}
			j_{a_2}  &  j_{a_1}  & \lambda \\
			J_{1}    &  J_2      & j_{b_1}
		\end{array}
		\right\} ,
\end{equation}

\begin{equation}
	\langle (a_2 b_2)_{J_2}|| Q_{\lambda} || (a_1 b_1)_{J_1} \rangle_{(2)}
	= \mathcal{N}_{a_1 b_1}^{J_1} \mathcal{N}_{a_2 b_2}^{J_2} 
	\hat{J_1} \hat{J}_2
	\delta_{a_1 b_2}
	u_{a_2} u_{b_1} 
	{ (-)^{j_{a_2} + j_{b_1} + \lambda}}
	\langle j_{a_2} ||Q_{\lambda}|| j_{b_1} \rangle
	\left\{
	\begin{array}{ccc}
		j_{a_2}  &  j_{b_1}  & \lambda \\
		J_{1}    &  J_2      & j_{a_1}
	\end{array}
	\right\} ,
\end{equation}

\begin{equation}
	\langle (a_2 b_2)_{J_2}|| Q_{\lambda} || (a_1 b_1)_{J_1} \rangle_{(3)}
	= \mathcal{N}_{a_1 b_1}^{J_1} \mathcal{N}_{a_2 b_2}^{J_2}
	\hat{J_1} \hat{J}_2
	\delta_{a_1 a_2}
	u_{b_2} u_{b_1} (-)^{j_{b_1} + j_{a_2} + J_2 + \lambda}
	\langle j_{b_2} ||Q_{\lambda}|| j_{b_1} \rangle
	\left\{
	\begin{array}{ccc}
		j_{b_2}  &  j_{b_1}  & \lambda \\
		J_{1}    &  J_2      & j_{a_1}
	\end{array}
	\right\} ,
\end{equation}

\begin{equation}
	\langle (a_2 b_2)_{J_2}|| Q_{\lambda} || (a_1 b_1)_{J_1} \rangle_{(4)}
	= \mathcal{N}_{a_1 b_1}^{J_1} \mathcal{N}_{a_2 b_2}^{J_2}
	\hat{J_1} \hat{J}_2
	\delta_{b_1 a_2}
	u_{b_2} u_{a_1} 
	{ (-)^{J_1 + J_2 + \lambda+1}}
	\langle j_{b_2} ||Q_{\lambda}|| j_{a_1} \rangle
	\left\{
	\begin{array}{ccc}
		j_{b_2}  &  j_{a_1}  & \lambda \\
		J_{1}    &  J_2      & j_{b_1}
	\end{array}
	\right\} ,
\end{equation}

\begin{equation}
	\langle (a_2 b_2)_{J_2}|| Q_{\lambda} || (a_1 b_1)_{J_1} \rangle_{(5)}
	= \mathcal{N}_{a_1 b_1}^{J_1} \mathcal{N}_{a_2 b_2}^{J_2}
	\hat{J}_1 \hat{J}_2 
	\delta_{b_1 b_2}
	v_{a_1} v_{a_2}
	\langle j_{a_1} ||Q_{\lambda}|| j_{a_2} \rangle
	(-)^{j_{a_1} + j_{b_1} + J_{1}+1}
	\left\{
	\begin{array}{ccc}
		j_{a_2}  &  j_{a_1}  & \lambda \\
		J_{1}    &  J_2      & j_{b_1}
	\end{array}
	\right\} ,
\end{equation}

\begin{equation}
	\langle (a_2 b_2)_{J_2}|| Q_{\lambda} || (a_1 b_1)_{J_1} \rangle_{(6)}
	= \mathcal{N}_{a_1 b_1}^{J_1} \mathcal{N}_{a_2 b_2}^{J_2}
	\hat{J_1} \hat{J}_2
	\delta_{a_1 b_2}
	v_{a_2} v_{b_1} 
	\langle j_{b_1} ||Q_{\lambda}|| j_{a_2} \rangle
	\left\{
	\begin{array}{ccc}
		j_{a_2}  &  j_{b_1}  & \lambda \\
		J_{1}    &  J_2      & j_{a_1}
	\end{array}
	\right\} ,
\end{equation}

\begin{equation}
	\langle (a_2 b_2)_{J_2}|| Q_{\lambda} || (a_1 b_1)_{J_1} \rangle_{(7)}
	= \mathcal{N}_{a_1 b_1}^{J_1} \mathcal{N}_{a_2 b_2}^{J_2}
	\hat{J_1} \hat{J}_2
	\delta_{a_1 a_2}
	v_{b_2} v_{b_1} (-)^{j_{a_2} + j_{b_2} + J_2 + 1}
	\langle j_{b_1} ||Q_{\lambda}|| j_{b_2} \rangle
	\left\{
	\begin{array}{ccc}
		j_{b_2}  &  j_{b_1}  & \lambda \\
		J_{1}    &  J_2      & j_{a_1}
	\end{array}
	\right\} ,
\end{equation}

\begin{equation}
	\langle (a_2 b_2)_{J_2}|| Q_{\lambda} || (a_1 b_1)_{J_1} \rangle_{(8)}
	= \mathcal{N}_{a_1 b_1}^{J_1} \mathcal{N}_{a_2 b_2}^{J_2}
	\hat{J_1} \hat{J}_2
	\delta_{b_1 a_2}
	v_{b_2} v_{a_1} 
	{ (-)^{j_{a_1} + j_{b_2} + J_1 + J_2+1}}
	\langle j_{a_1} ||Q_{\lambda}|| j_{b_2} \rangle
	\left\{
	\begin{array}{ccc}
		j_{b_2}  &  j_{a_1}  & \lambda \\
		J_{1}    &  J_2      & j_{b_1}
	\end{array}
	\right\} ,
\end{equation}
where $\mathcal{N}_{a b}^{J}$ is a normalization factor,
\begin{equation}
\mathcal{N}_{a b}^{J} = \frac{\sqrt{1+(-)^{J}\delta_{ab}}}{1+\delta_{ab}}.
\end{equation}

Diagrams A  and C corresponds to 
$\langle (a_2 b_2)_{J_2}|| Q_{\lambda} || (a_1 b_1)_{J_1} \rangle_{[(1) + (5)]}$, where there is a $\delta_{b_1 b_2}$.
Diagrams A$'$ and C$'$ corresponds to 
$\langle (a_2 b_2)_{J_2}|| Q_{\lambda} || (a_1 b_1)_{J_1} \rangle_{[(2) + (6)]}$, where there is a $\delta_{a_1 b_2}$.
Diagrams B  and D corresponds to 
$\langle (a_2 b_2)_{J_2}|| Q_{\lambda} || (a_1 b_1)_{J_1} \rangle_{[(3) + (7)]}$, where there is a $\delta_{a_1 a_2}$. 
Diagrams B$'$ and D$'$ corresponds to 
$\langle (a_2 b_2)_{J_2}|| Q_{\lambda} || (a_1 b_1)_{J_1} \rangle_{[(4) + (8)]}$, where there is a $\delta_{b_1 a_2}$.

The other two terms $\alpha_{\bar{a}} \alpha_{\bar{b}}$ and $\alpha^{\dag}_{a} \alpha^{\dag}_{b}$  in $Q_{\lambda\mu}$
lead to the quasiparticle-pair annihilation and creation,
\begin{equation}
\langle 0|| Q_{\lambda} || (ab)_{\lambda} \rangle
=\mathcal{N}_{a b}^{\lambda}   [
(-)^{j_{a} + j_{b} +\lambda } 
v_{b} u_{a} \langle j_b ||Q_{\lambda}|| j_a \rangle 
- v_{a} u_{b}  
\langle j_a ||Q_{\lambda}|| j_b \rangle ],
\end{equation}
\begin{equation}
\langle (ab)_{\lambda} || Q_{\lambda} || 0\rangle
=\mathcal{N}_{a b}^{\lambda}  [
- \langle j_a || Q_{\lambda} || j_b \rangle u_{a} v_{b}
+ (-)^{j_a + j_{b} + \lambda}
\langle j_b || Q_{\lambda} || j_a \rangle u_{b} v_{a} ],
\end{equation}
which are involved in diagrams E to H$'$ and diagrams I to L$'$ respectively.

\section{QPVC matrix elements}
\label{Apdx_qpvc_me}

\begin{figure*}[tbph]
	\centering
	\includegraphics[width=0.8\textwidth]{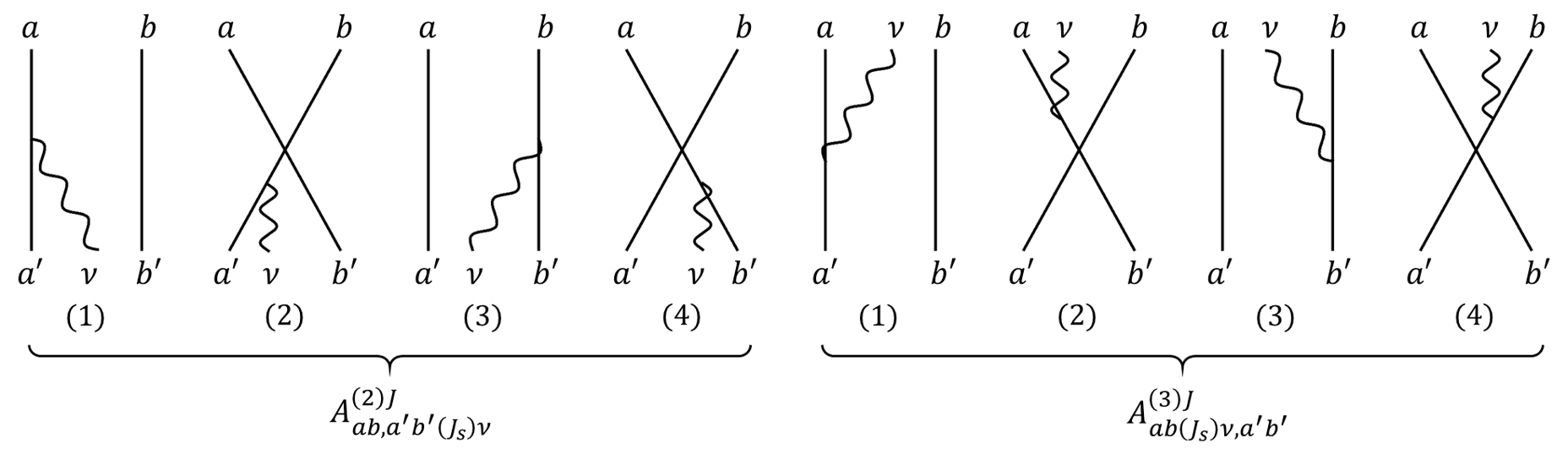}
	\caption{Diagrams for the QPVC matrix elements $A^{(2)J}_{ab,a' b' (J_s) \nu}$ and $A^{(3)J}_{ab(J_s)\nu,a' b'}$.}
	\label{fig:figA1}
\end{figure*}

The QPVC matrix elements $A^{(2)}$ and $A^{(3)}$ are 
\begin{equation}
A^{(2)JM}_{ab,a'b'(J_{s})\nu} = 
\langle 0 | \left[ A_{ab}(JM), [ V, B^{\dag}_{a'b'(J_{s})\nu}(JM) ] \right] |0 \rangle,
\end{equation}
\begin{equation}
A^{(3)JM}_{ab(J_{s})\nu,a'b'}   =
 \langle 0 |\left[ B_{ab(J_{s})\nu}(JM), [ V, A^{\dag}_{a'b'}(JM) ] \right] |0 \rangle,
\end{equation}
where $A^{\dag}$ creates a quasiparticle-pair state (2qp), and $B^{\dag}$ creates a 2qp$\otimes$phonon state,
\begin{equation}
A^{\dag}_{ab}(JM)
= \mathcal{N}_{ab}^{J} \sum_{m_a m_b} C_{j_a m_a j_b m_b}^{JM}
\alpha^{\dag}_{a m_a} \alpha^{\dag}_{b m_b},
\end{equation}
\begin{equation}
B^{\dag}_{ab(J_s)\nu}(JM)
=  \sum_{M_{s} \mu} C_{J_{s} M_{s} \lambda \mu}^{JM}
A^{\dag}_{ab}(J_{s} M_{s}) Q^{\dag}_{\nu\lambda\mu} .
\end{equation}
$Q^{\dag}_{\nu\lambda\mu}$ is the phonon creation operator in the QRPA model,
\begin{equation}
 Q^{\dag}_{\nu\lambda\mu} =
\sum_{ab} X^{\nu\lambda}_{ab} A^{\dag}_{ab}(\lambda\mu)
        - (-)^{\lambda+\mu} Y^{\nu\lambda}_{ab} A_{ab}(\lambda-\mu).
\end{equation}

The expressions of the QPVC matrix elements $A^{(2)}$ are
\begin{subequations}
	\begin{align}
		A^{(2) J}_{ab,a' b'(J_{s}) \nu} (1)
		=& \delta_{bb'} 	
		\mathcal{N}_{ab}^{J}
		\mathcal{N}_{a'b'}^{J_{s}}
		(-)^{ j_{a} + j_{b'} + \lambda + J_{s}}
		\hat{J}_{s} 
		\left\{
		\begin{array}{ccc}
			j_{a'}  &  j_{b'}  &  J_{s} \\
			J       &  \lambda'&  j_{a}
		\end{array}
		\right\} 
		\langle a|| V ||a' \nu \rangle ; \\
	%
	A^{(2) J}_{ab,a' b'(J_{s}) \nu} (2)
	=& (-)\delta_{ab'} 
	\mathcal{N}_{ab}^{J}
	\mathcal{N}_{a'b'}^{J_{s}}
	(-)^{j_{b} + j_{b'} + \lambda + J_{s}} 
	\hat{J}_{s}
	\left\{
	\begin{array}{ccc}
		j_{a'}   & j_{b'}   & J_{s} \\
		J        & \lambda & j_{b}
	\end{array}
	\right\}  (-)^{j_a + j_b + J}
	\langle b || V || a' \nu \rangle ; \\
	%
		A^{(2) J}_{ab,a' b'(J_{s}) \nu} (3)
		=& \delta_{aa'} 
		\mathcal{N}_{ab}^{J}
		\mathcal{N}_{a'b'}^{J_{s}}
		(-)^{j_{a'} + j_{b'} + \lambda + J}
		\hat{J}_{s}
		\left\{
		\begin{array}{ccc}
			j_{a'}   & j_{b'} & J_{s} \\
			\lambda & J      & j_{b}
		\end{array}
		\right\} 
		\langle b|| V || b' \nu \rangle ; \\
	%
	A^{(2) J}_{ab,a' b'(J_{s}) \nu} (4)
	=& 
	(-)\delta_{ba'}
	\mathcal{N}_{ab}^{J}
	\mathcal{N}_{a'b'}^{J_{s}}
	(-)^{j_{a'} + j_{b'} + \lambda + J}
	\hat{J}_{s}
	\left\{
	\begin{array}{ccc}
		j_{a'}   & j_{b'} & J_{s} \\
		\lambda & J      & j_{a}
	\end{array}
	\right\}
	(-)^{j_a + j_b + J} 
	\langle a || V || b' \nu \rangle .
	\end{align}
\end{subequations}
$A^{(3)}$ can be easily obtained by the following relation,
\begin{equation}
  A^{(3) J}_{a' b'(J_{s}) \nu, ab}(k) = [A^{(2) J}_{ab, a' b'(J_{s}) \nu} (k) ]^{\ast},
  ~~ k = 1, 2, 3, 4 
\end{equation}
where $\langle a|| V ||b\nu\rangle$ is the QPVC vertex.
Because the QPVC matrix elements $A^{(2)}$ and $A^{(3)}$ are independent from the magnetic quantum number $M$, 
we omit it throughout the paper.

For diagrams E, E$'$, F, and F$'$, the QPVC matrix elements 
$A^{(3) J_i}_{a_2 b_2 (\lambda) N_f, a_1 b_1}(1)$, $A^{(3) J_i}_{a_2 b_2 (\lambda) N_f, a_1 b_1}(2)$, 
$A^{(3) J_i}_{a_2 b_2 (\lambda) N_f, a_1 b_1}(3)$, and $A^{(3) J_i}_{a_2 b_2 (\lambda) N_f, a_1 b_1}(4)$
are involved. 
For diagrams G, G$'$, H, and H$'$, the QPVC matrix elements 
$A^{(2) \lambda}_{a_2 b_2, a_1 b_1(J_f) N_i}(1)$, $A^{(2) \lambda}_{a_2 b_2, a_1 b_1(J_f) N_i}(2)$, 
$A^{(2) \lambda}_{a_2 b_2, a_1 b_1(J_f) N_i}(3)$, and $A^{(2) \lambda}_{a_2 b_2, a_1 b_1(J_f) N_i}(4)$
are involved. 
For diagrams I, I$'$, J, and J$'$, the QPVC matrix elements 
$A^{(2) J_f}_{a_2 b_2, a_1 b_1 (\lambda) N_i }(1)$, $A^{(2) J_f}_{a_2 b_2, a_1 b_1 (\lambda) N_i }(2)$, 
$A^{(2) J_f}_{a_2 b_2, a_1 b_1 (\lambda) N_i }(3)$, and $A^{(2) J_f}_{a_2 b_2, a_1 b_1 (\lambda) N_i }(4)$
are involved. 
For diagrams K, K$'$, L, and L$'$, the QPVC matrix elements 
$A^{(3) \lambda}_{a_2 b_2(J_i)N_f, a_1 b_1}(1)$, $A^{(3) \lambda}_{a_2 b_2(J_i)N_f, a_1 b_1}(2)$, 
$A^{(3) \lambda}_{a_2 b_2(J_i)N_f, a_1 b_1}(3)$, and $A^{(3) \lambda}_{a_2 b_2(J_i)N_f, a_1 b_1}(4)$
are involved. 

\section{QPVC and QRPA vertices}
\label{Apdx_vertex}
The QPVC vertex is evaluated as \cite{Niu2016,Li2024}
\begin{equation}
	\begin{aligned}
		\langle a|| V || a'\nu'  \rangle
		=& 
		\hat{\lambda}'  \mathcal{N}_{aa'}^{\lambda' -1}
		\sum_{a_0 b_0} 
		X^{\nu'\lambda'}_{a_0 b_0}
		[( v_{a}^{\ast} v_{a'} v_{a_0} u_{b_0} - u_{a}^{\ast} u_{a'} u_{a_0} v_{b_0} )
		\langle [a (a')^{-1}]_{\lambda'} | V | [a_0 (b_0)^{-1}]_{\lambda'} \rangle  \\ & \qquad \qquad
		+(-)^{j_{a_0} + j_{b_0}+\lambda'}
		( u_{a}^{\ast} u_{a'} v_{a_0} u_{b_0} - v_{a}^{\ast} v_{a'} u_{a_0} v_{b_0} )
		\langle [a (a')^{-1}]_{\lambda'} | V | [b_0 (a_0)^{-1}]_{\lambda'} \rangle  \\ & \qquad \qquad
		-( v_{a}^{\ast} u_{a'} v_{a_0} v_{b_0} - u_{a}^{\ast} v_{a'} u_{a_0} u_{b_0} ) 
		\langle (a a')_{\lambda'} | V | (a_0 b_0)_{\lambda'} \rangle ]
		\\& 
		- (-)^{j_{a}+j_{a'}+\lambda'}
		Y^{\nu'\lambda'}_{a_0 b_0} 
		[(v_{a'} v_{a}^{\ast} v_{a_0}^{\ast} u_{b_0}^{\ast} 
		-u_{a'}^{\ast} u_{a}^{\ast} u_{a_0}^{\ast} v_{b_0} ) 
		\langle [a' (a)^{-1}]_{\lambda'}|V| [a_{0} (b_0)^{-1}]_{\lambda'} \rangle  \\ & \qquad \qquad
		+(-)^{j_{a_0}+j_{b_0}+\lambda'}
		(u_{a'}^{\ast} u_{a}^{\ast} v_{a_0} u_{b_0}^{\ast}
		-v_{a'} v_{a}^{\ast} u_{a_0}^{\ast} v_{b_0}^{\ast} ) 
		\langle [a' (a)^{-1}]_{\lambda'}|V| [b_{0} (a_0)^{-1}]_{\lambda'} \rangle  \\ & \qquad \qquad
		-(v_{a'} u_{a}^{\ast} v_{a_0}^{\ast} v_{b_0}^{\ast}
		-u_{a'}^{\ast} v_{a} u_{a_0}^{\ast} u_{b_0}^{\ast} ) 
		\langle (a' a)_{\lambda'} | V | (a_0 b_0)_{\lambda'} \rangle ] ,
	\end{aligned}
\end{equation}
with the coupled two-body interaction matrix elements
\begin{equation}
  \langle a (a')^{-1} ; \lambda' | V | a_0 (b_0)^{-1}; \lambda' \rangle
= \mathcal{N}_{aa'}^{\lambda'} \mathcal{N}_{a_0 b_0}^{\lambda'}
  \sum_{m_a m_{a'} m_{a_0} m_{b_0}}
  (-)^{j_{a'} - m_{a'}+j_{b_0}-m_{b_0}} 
  C_{j_a m_a j_{a'} -m_{a'}}^{\lambda' \mu'}
  C_{j_{a_0} m_{a_0} j_{b_0} -m_{b_0}}^{\lambda' \mu'}
  \bar{v}_{{a} {b}_0 {a}' {a_0}} ,
\end{equation}
\begin{equation}
  \langle a a' ;\lambda' | V | a_0 b_0 ;\lambda' \rangle
= \mathcal{N}_{aa'}^{\lambda'} \mathcal{N}_{a_0 b_0}^{\lambda'}
  \sum_{m_a m_{a'} m_{a_0} m_{b_0}} 
  C_{j_{a} m_{a} j_{a'} m_{a'}}^{\lambda' \mu'}
  C_{j_{a_0} m_{a_0} j_{b_0} m_{b_0}}^{\lambda' \mu'}
  \bar{v}_{{a} {a}' {a}_0 {b}_0} . 
\end{equation}

The QRPA vertices $\langle \nu || V || a b \rangle$ and $\langle ab,\nu || V || 0 \rangle$
can be easily evaluated by $X_{ab}^{\nu\lambda}$ and $Y_{ab}^{\nu\lambda}$ in the QRPA model,
\begin{equation}
\langle \nu || V || a b \rangle
= \hat{\lambda} [\Omega_{\nu} - (E_{a} + E_{b})] X_{ab}^{\nu\lambda}, 
\end{equation}
\begin{equation}
\langle ab,\nu || V || 0 \rangle
= -\hat{\lambda} [\Omega_{\nu} + (E_{a} + E_{b})] Y_{ab}^{\nu\lambda}.
\end{equation}
\end{widetext}

%

\end{document}